\documentclass{PoS}

\title{Sterile neutrinos along the DUNE decay pipe}

\ShortTitle{Sterile $\nu$'s in DUNE}

\author{Jo\~{a}o Penedo\\
        Centro de F\'{\i}sica Te\'{o}rica de Part\'{\i}culas, Instituto Superior T\'{e}cnico,
        Av. Rovisco Pais, 1049-001, Lisboa, Portugal\\
	E-mail: \email{joao.t.n.penedo@tecnico.ulisboa.pt}}

\author{\speaker{Jo\~{a}o Pulido}\\
        Centro de F\'{\i}sica Te\'{o}rica de Part\'{\i}culas, Instituto Superior T\'{e}cnico,
        Av. Rovisco Pais, 1049-001, Lisboa, Portugal\\
        E-mail: \email{pulido@cftp.ist.utl.pt}}

\abstract{We analyse the sensitivity of the Deep Underground Neutrino Experiment (DUNE) to a sterile 
         neutrino, combining information from both the Near Detector (ND) and the Far Detector
         (FD). DUNE's sterile exclusion reach is affected by taking into account the information on 
	 the neutrino production point, in contrast to assuming a point-like neutrino source.
         Visible differences remain after taking into account energy bin-to-bin uncorrelated 
	 systematics.}

\FullConference{8th Symposium on Prospects in the Physics of Discrete Symmetries\\
		DISCRETE 2022\\
		7-11 November 2022\\
                Kongresshaus Baden-Baden, Germany}

\begin{document}

%\section{Introduction}

The first hint for sterile neutrinos came from the LSND experiment in which an intense proton beam
hitting a target ultimately produced $\nu_{\mu},~\bar\nu_{\mu},~\nu_e$ and $e^{+}$ from $\pi^{+} 
\rightarrow \mu^{+}\nu_{\mu}$ and $\mu^{+} \rightarrow e^{+} \nu_e\bar\nu_{\mu}$ decays  
\cite{LSND:1996ubh}. The unexpected observation of $\bar\nu_e$'s ($\bar\nu_e$ appearance) 
yielded the hypothesis that some of the $\bar\nu_{\mu}$'s could oscillate into $\bar\nu_{e}$'s
through sterile neutrinos. More recently, it has been noted that sterile neutrinos also provide 
a viable explanation for the Gallium and reactor antineutrino anomalies (for a review see 
\cite{Giunti:2019aiy}). For the first of these, artificial neutrino radioactive sources were placed 
inside the GALLEX \cite{Kaether:2010ag} and SAGE \cite{Abdurashitov:2005tb} Ga detectors for 
calibration in order to measure the reaction $\nu_e + ^{71}{\rm Ga} \rightarrow e^{-}+^{71}{\rm Ge}$. 
A deficit of the observed rate with respect to the well measured activity of the sources was found: 
$\bar R=0.844\pm 0.031$ for the dominant Ge ground state production mode, with the contributions to 
its excited states being very similar \cite{Giunti:2022btk}. For the reactor antineutrino 
anomaly (RAA) \cite{Mention:2011rk}, fewer than expected antineutrinos from radioactive nuclides in 
the reaction $\bar\nu_e + p \rightarrow e^{+} + n$ were found, with $\bar R=0.936\pm^{0.024}_{0.023}$
\cite{Giunti:2021kab}. 
A possible solution to this $\nu_e$ $(\bar\nu_e)$ disappearance is to introduce one sterile neutrino 
(3+1 scenario) with
\begin{equation}
\Delta m_{41}^2\simeq\Delta m_{42}^2\simeq\Delta m_{43}^2=O(1~{\rm eV}^2)~,~\sin^2\theta_{14}\sim 0.1.
\end{equation}
However, upon this assumption, a larger than $2\sigma$ tension arises between the Ga and RAA 
preferred values \cite{Giunti:2022btk}
$$\Delta m^2_{41}({\rm Ga})\sim 2\Delta m^2_{41}(\rm Reactor)~~~~sin^2 2\theta_{14}(Ga)
\sim 3 sin^2 2\theta_{14}(\rm Reactor).$$
Given the order of magnitude of $\Delta m^2_{41}$, the possible active/sterile oscillations are 
short-baseline (SBL) and, if they exist, they must show up in $\nu_{\mu}$, $\bar\nu_{\mu}$
disappearance which has not been the case so far. On the other hand, in the approximation of
small mixing, the following relation holds
$$\sin^2 2\theta_{e\mu}\simeq \frac{1}{4}\sin^2 2\theta_{ee} \sin^2 2\theta_{\mu\mu}$$
(with $\sin^2 2\theta_{e\mu}=\sin^2 2\theta_{14}\sin^2 \theta_{24},~\theta_{ee}=\theta_{14},
~\theta_{\mu\mu}\simeq\theta_{24}).$ However, evidence from several experiments shows that \footnote
{This list of experiments is by all means incomplete.}
\begin{itemize}
\item $\sin^2 2\theta_{e\mu}\gtrsim 10^{-3}$ from $\nu_e$, $\bar\nu_e$ appearance (e. g. LSND, Karmen 
\cite{KARMEN:2002zcm})
\item $\sin^2 2\theta_{ee}\sim (0.15-0.2)$ from $\nu_e$, $\bar\nu_e$ disappearance, however in tension 
(Ga, RAA)
\item $\sin^2 2\theta_{\mu\mu}\lesssim 10^{-2}$ from $\nu_{\mu}$, $\bar\nu_{\mu}$ disappearance (e. g. 
MINOS\&MINOS+ \cite{MINOS:2017cae}).
\end{itemize}	
Moreover, recent constraints on $\sin^2 2\theta_{e\mu}$ slightly contradict LSND 
\cite{MINOS:2020iqj}
$$\sin^2 2\theta_{e\mu} < 10^{-3}$$
unless  $\Delta m^2_{41} > 3~{\rm eV}^2$, which in turn implies a tension with Ga and reactor 
antineutrino anomalies. Hence the sterile neutrino issue remains unsettled. 

To this end, the Deep Underground Neutrino Experiment (DUNE) setup \cite{DUNE:2021cuw}, whose nominal 
mission is to perform precise measurements of neutrino properties and oscillations, may play an 
important clarifying role. In this experiment, neutrinos are produced from meson decays along a decay 
pipe originating from proton collisions on a graphite target (see fig.1), so the neutrino source is 
smeared in space rather than point-like. In the 3+0 case (no steriles), owing to the smallness of 
$\Delta m^2$'s, no oscillations can occur up to the near detector (ND). On the other hand, in the 
3+1 case active/sterile oscillations can take place upstream from the ND, since the order of 
magnitude of $\Delta m^2_{41}$ $(0.1-1~{\rm eV}^2)$ implies this oscillation length to be of the 
order of the decay pipe length.

Given the fact that the neutrino origin is distributed along the decay pipe (fig.2) and oscillation 
to steriles occurs prior to the ND, a point-source approximation would be erroneous and 
source-volume effects need to be explicitly taken into account. The objective of our present work 
is thus to evaluate the effect of sterile neutrinos in DUNE ND and FD event rates from a statistical 
analysis, assuming a neutrino smeared source. 

\begin{figure}[htb]
\centering
\includegraphics[height=20mm,width=\textwidth]{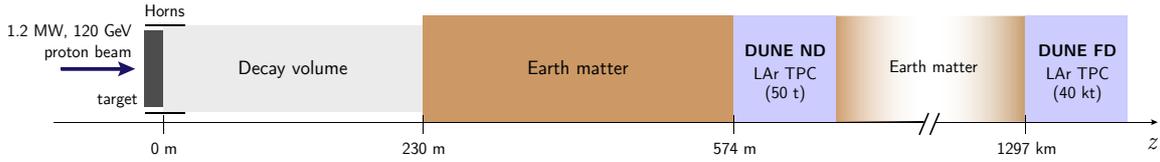}
\caption{Diagram of the DUNE beam setup, where $z$ represents the distance from the graphite target.} 
\label{fig.1}
\end{figure}

We used the GLoBES software \cite{Huber:2004ka, Huber:2007ji} and conceptually divided the 
decay pipe into 30 sections, associating each to a different point-source with its fixed 
baseline $L$. The flux arriving at the ND from each section is passed to GLoBES as the flux of an 
independent experiment \footnote{GLoBES internal functions are modified so that one effectively 
works with two experiments/detectors when computing $\chi^2$.}.

%\vspace{-0.25cm}[htb]
\begin{figure}[htb]
\centering
\includegraphics[height=70mm,width=90mm]{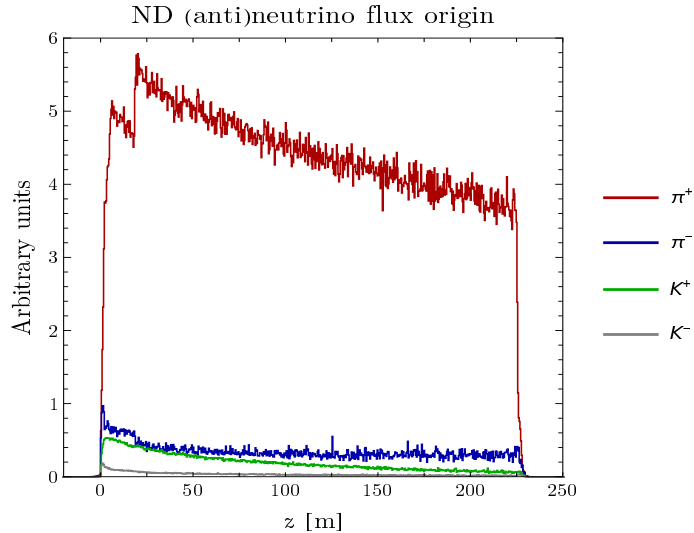}
\caption{The origin of the neutrinos and antineutrinos from $\pi$ and $K$ decays reaching the ND.}
\end{figure}

In the 3+1 framework, the propagation of neutrinos in matter is described by the Hamiltonian ($\alpha,
\beta = e,\mu,\tau,s$)
\vspace{-0.4cm}
\begin{equation}
{H^{\rm mat}_{\alpha\beta}\!=\!\frac{1}{2E}\!\left[U~{\rm diag}(0,\Delta m^2_{21},\Delta m^2_{31},
\Delta m^2_{41})U^{\dagger}+{\rm diag}(A_{CC},0,0,A_{NC})\right]}
\label{H_mat}
\end{equation}
where $U$ is the Pontecorvo-Maki-Nakagawa-Sakata (PMNS) matrix, $A_{CC}=2\sqrt{2}G_F N_e E,
~A_{NC}=\sqrt{2}G_F N_n E$ and $N_e$, $N_n$ are the electron and neutron densities in the medium
($N_e\simeq N_n$). The matter Hamiltonian (\ref{H_mat}) can be brought, apart from a constant term
which does not affect the oscillation probabilities, to a vacuum form (identical to the first 
term in (\ref{H_mat})) with the replacements $U\rightarrow \tilde U$, $\Delta m^2_{ij}\rightarrow
\Delta\tilde m^2_{ij}$ where the tildes denote the values of these quantities in matter. Denoting
the eigenvalues of the matrix $2EH^{\rm mat}$ by $\hat\Delta m^2_{i1}~(i=1,...,4)$, eq.(\ref{H_mat}) 
becomes (for the analytical formulation see \cite{Penedo:2022etl}),
\begin{equation}
{H^{\rm mat}_{\alpha\beta}\!=\!\frac{1}{2E}\!\left[U~{\rm diag}(0,\Delta\tilde m^2_{21},\Delta\tilde
	m^2_{31},\Delta\tilde m^2_{41})U^{\dagger}+\hat\Delta m^2_{11}{\rm diag}(1,1,1,1) \right]}.
\label{H_2}
\end{equation}

At the ND, the oscillation probability (short-baseline) depends on the point of production of the 
(active) $\alpha$ flavour neutrino, located at a distance $L=L_1+L_2$. We have \cite{Penedo:2022etl} 

\begin{equation}
P^{SBL}_{\alpha\beta}(L_i,E)=\left|\sum_{\gamma}\sum_{j,k}\tilde U_{\beta j}\tilde U^{\ast}_{\gamma j} 
U_{\gamma k}U^{\ast}_{\alpha k} \exp\left(-i\frac{\Delta m^2_{k1}L_1 +\Delta\tilde m^2_{j1}L_2}{2E}
\right)\right|^2
\label{SBL}
\end{equation}
where $L_1$ and $L_2$ denote the distances travelled by the neutrinos in vacuum and in matter up
to the ND.

For both ND and FD event rates, a low-pass filter must be applied at the probability level to
appropriately average out unresolvable fast oscillations. At the FD, the rates are expected to 
be sensitive simply to the matter density, $\rho\simeq 2.6~{\rm g~cm}^{-3}$, so only the $\tilde U$'s 
and the $\Delta\tilde m^2$'s remain in the long baseline (LBL) probability expression
\begin{equation}
P^{LBL}_{\alpha\beta}(L,E)=\sum_{j,j~'}\tilde U_{\beta j}\tilde U^{\ast}_{\alpha j}\tilde U_{\alpha j~'}
\tilde U^{\ast}_{\beta j~'}\exp\left(-i\frac{\Delta\tilde m^2_{j j~'}L}{2E}\right)\exp\left[
-\frac{\sigma^2_E}{2E^2}\left(\frac{\Delta\tilde m^2_{j j~'}L}{2E}\right)^2\right]
\end{equation}
for a single baseline $L$ in matter. Here the rightmost exponential is a Gaussian low-pass filter 
averaging out the fast oscillations \cite{Huber:2020}. 

For the ND event rate, matter effects are negligible and it is enough to consider the following 
approximate result obtained in the limit of vanishing matter density and vanishing standard neutrino
mass squared differences \cite{Penedo:2022etl}. 
%oscillation to active neutrinos do not have 
%time to occur, so that the following approximate expression derived from eq.(\ref{SBL}) is sufficiently 
%accurate \cite{Penedo:2022etl}. We have, with the Gaussian filter included
\begin{equation}
P^{SBL}_{\alpha\beta}(L,E) \simeq \delta_{\alpha\beta}-2|U_{\alpha 4}|^2\left(\delta_{\alpha\beta}-
|U_{\beta 4}|^2\right)\left\{1-\cos\left(\frac{\Delta m^2_{41}L}{2E}\right)\exp\left[
-\frac{\sigma^2_E}{2E^2}\left(\frac{\Delta m^2_{41}L}{2E}\right)^2\right]\right\}. 
\end{equation}
In the 3+0 case at the ND and in forward horn current (FHC) mode, where the dominant process is
$\pi^{+}\rightarrow \mu^{+}\nu_{\mu}$, the signal event rate is composed solely of $\nu_{\mu}$, 
whereas in reverse horn current (RHC) mode, with $\pi^{-}\rightarrow \mu^{-}\bar\nu_{\mu}$ as the dominant 
process, it is solely $\bar\nu_{\mu}$. In the 3+0 case at the FD and in FHC the signal is $\nu_{\mu}$ 
and $\nu_{e}$ (due to oscillations) and correspondingly in RHC it is $\bar\nu_{\mu}$ and $\bar\nu_{e}$. 

However, background from contamination ($\nu_e$, $\bar\nu_e$) and misisdentifications ($\nu_{\mu}$, 
$\bar\nu_{\mu}$ misidentified as $\nu_e$, $\bar\nu_e$) are present in both detectors and operation 
modes (FHC and RHC), and are therefore part of the event rates.

We consider the 3+1 case and follow the DUNE simulation configurations. For the statistical analysis,
the concept of `channel' and `rule' are essential. A `channel' consists of the physical oscillation process, 
the energy reconstruction and the detection process (neutral or charged current)\footnote{In our analysis we 
restrict ourselves to charged current (CC) processes.}, whereas a `rule' 
consists of a number of signal and background channels and their associated systematical uncertainties. 
Rules constitute the final link between the event rate and the statistical analysis. Hence, schematically
\begin{center}
Rule (event rate $\cup$ stat. analysis): signal, bkg, syst. uncertainties
\end{center}
For each of the two operation modes (FHC and RHC), two rules are to be considered: for FHC the search
is driven to neutrinos, while for RHC it is driven to antineutrinos. Hence there will be four rules altogether
%Each of the two operation modes unfolds in two rules, thus there are four rules to be considered
\begin{itemize}
\item FHC whose signal is $\nu_e$ (at the ND this signal is only possible in the 3+1 case)
\item RHC whose signal is $\bar\nu_e$ (at the ND this signal is only possible in the 3+1 case)
\item FHC whose signal is $\nu_{\mu}$
\item RHC whose signal is $\bar\nu_{\mu}$
\end{itemize}
All other channels in each of these rules are considered background. Note however that rules are
not 100\% efficient: some $\bar\nu$'s  and $\nu$'s contaminate FHC and RHC respectively. 

We use the following definition of $\chi^2$
\vspace{-0.3cm}
\begin{equation}
\chi^2=\chi^2_{\rm stat}(\omega,\omega_0,\zeta,\zeta')+\chi^2_{\rm prior}(\omega,\omega_0)+
\sum_{k=1}^{26}\left(\frac{\zeta_k}{\sigma_k}\right)^2+\sum_{r=1}^{4}\sum_{i=1}^{60}
\left(\frac{\zeta'_{r,i}}{\sigma'}\right)^2 
\label{chi2}
\end{equation}
for 60 energy bins ($i$) with energy $E_i \in[1.25,18.0]$ GeV in each rule ($r$). Two different
possibilities were investigated: one in which energy bin errors are neglected and one with bin-to-bin
uncorrelated shape uncertainties $\zeta{'}\!_{r,i}$ with $\sigma^{'}=5\%$. In eq.(\ref{chi2}), $\omega$
denotes the 7 measured random parameters $\Delta m^2_{21},\Delta m^2_{32},\theta_{12},\theta_{13},
\theta_{23},\rho_{ND},\rho_{FD}$ with central values $\omega_0$; $\zeta_k$ are the 26 normalization 
random parameters (ND and FD fiducial volumes, fluxes, cross sections) with their respective standard
deviations $\sigma_k$ given in table 3 of ref. \cite{DUNE:2020fgq}\footnote{See also table 1 of
\cite{Penedo:2022etl}.}. The statistical analysis proceeds
with either ($\Delta m^2_{41},\theta_{14}$) or ($\Delta m^2_{41},\theta_{24}$) fixed, marginalizing
respectively over $\theta_{24}$ or $\theta_{14}$ totally unconstrained along with all other 
parameters, namely $\theta_{34},\delta_{14},\delta_{24}$, also unconstrained. We have therefore 
7+26+1+3+240=277 marginalized random parameters. The $\chi^2$ minimization appears to be insensitive 
to the value of $\delta_{CP}$, which we fix at $1.28\pi$. The statistical and prior $\chi^2$ 
contributions in eq.(\ref{chi2}) are given by
%\vspace{-0.3cm}
\begin{equation}
\chi^2_{\rm stat}=2\sum_{d=1}^{2}\sum_{r=1}^{4}\sum_{i=1}^{60}\left[T^{d}_{r,i}-O^{d}_{r,i}
\left(1-\ln\frac{O^{d}_{r,i}}{T^{d}_{r,i}}\right)\right]~~,~~
%\end{equation}
%\begin{equation}
\chi^2_{\rm prior}=\sum_{j=1}^{7}\left(\frac{\omega_j-(\omega_0)_j}{\sigma(\omega_j)}\right)^2 
\end{equation}
%\vspace{-0.3cm}
with the test and observed event rates for each detector $d$ 
%\vspace{-0.3cm}
\begin{equation}
T^d_{r,i}=\sum_{\rm c=s,b}N^d_{r,c,i}(\omega)\left(1+\zeta'_{r,c,i}+\sum_{(k)}\zeta\right)~~,~~
O^d_{r,i}=\sum_{\rm c=s,b}N^d_{r,c,i}(\omega_0)
\end{equation}
\vspace{-0.3cm}\noindent
where the subscript $c=s,b$ denotes signal and background and $\displaystyle \sum_{(k)} \zeta$
is restricted to those $\zeta_k$ parameters involved in $d,c,r$.

Searching for $\chi^2_{min}=4.61$ (2 d.o.f.), we get the sterile neutrino exclusion plots (figs.3,4).

\begin{figure}[htb]
\centering
\includegraphics[height=65mm,width=58mm]{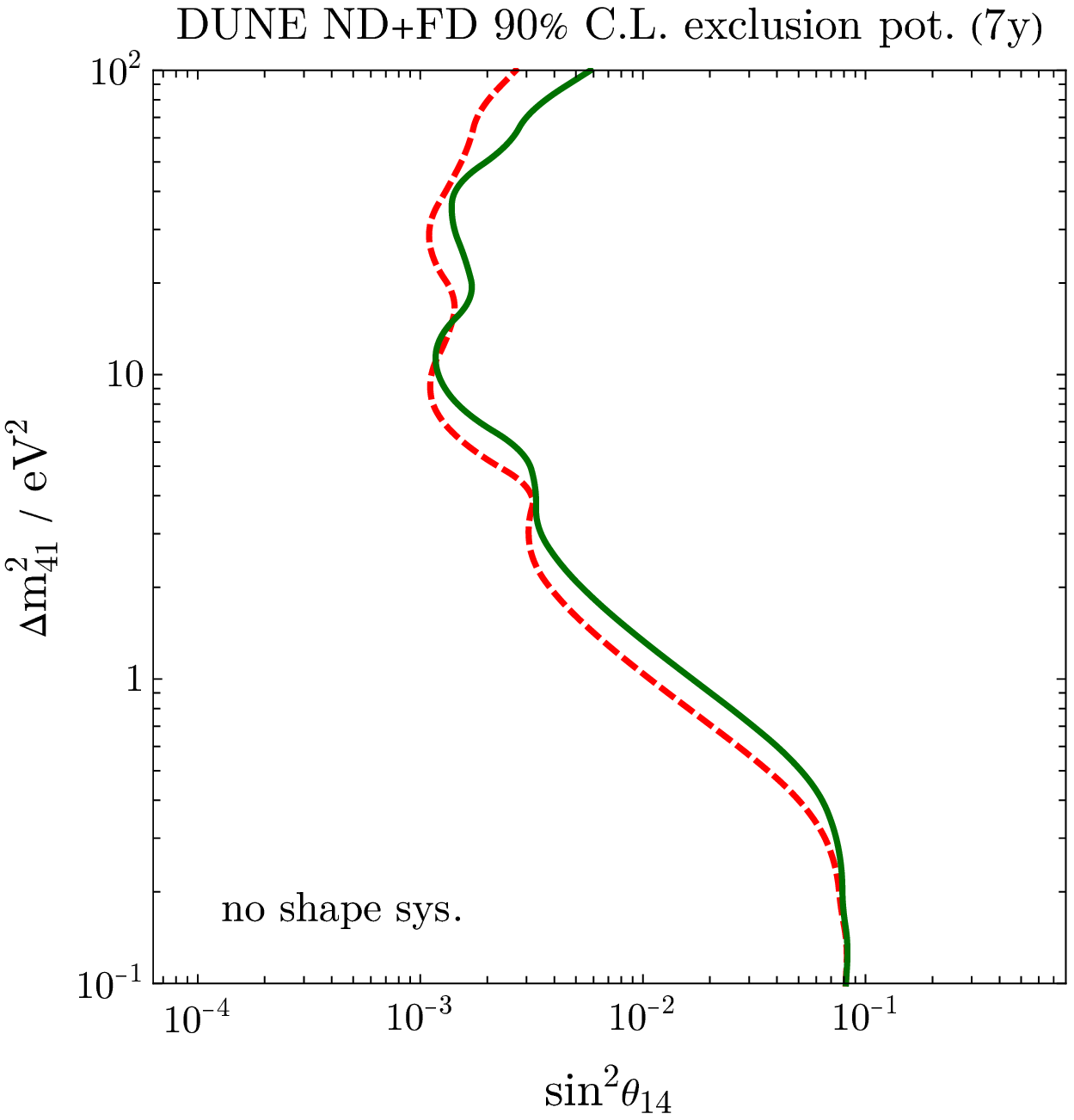}\quad
\includegraphics[height=65mm,width=58mm]{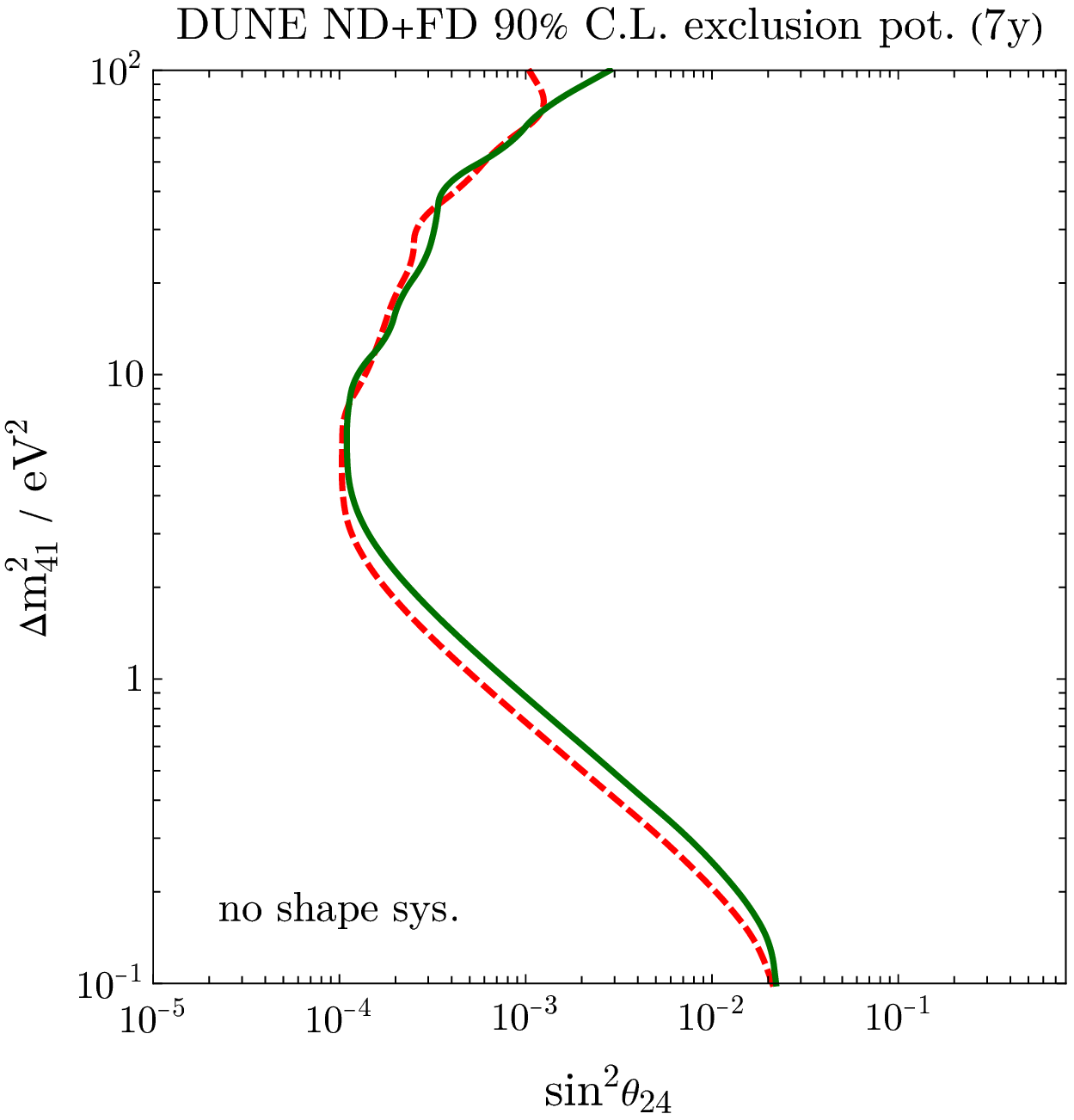}
\caption{The sterile exclusion potential of the combined DUNE ND and FD CC analyses at 90\% CL after
7 years of operation with no $\zeta'$ systematics. The dashed red curves are obtained assuming a 
common baseline $L=574$ m for ND oscillations and the solid green curves consider the smeared-source
effect.}
%oscillations while for the solid green curves the described smeared-source effect 
%Shape systematics are switched off, $\sigma'=0$}
\end{figure}

\vspace{0.5cm}

\begin{figure}[htb]
\centering
\includegraphics[height=65mm,width=58mm]{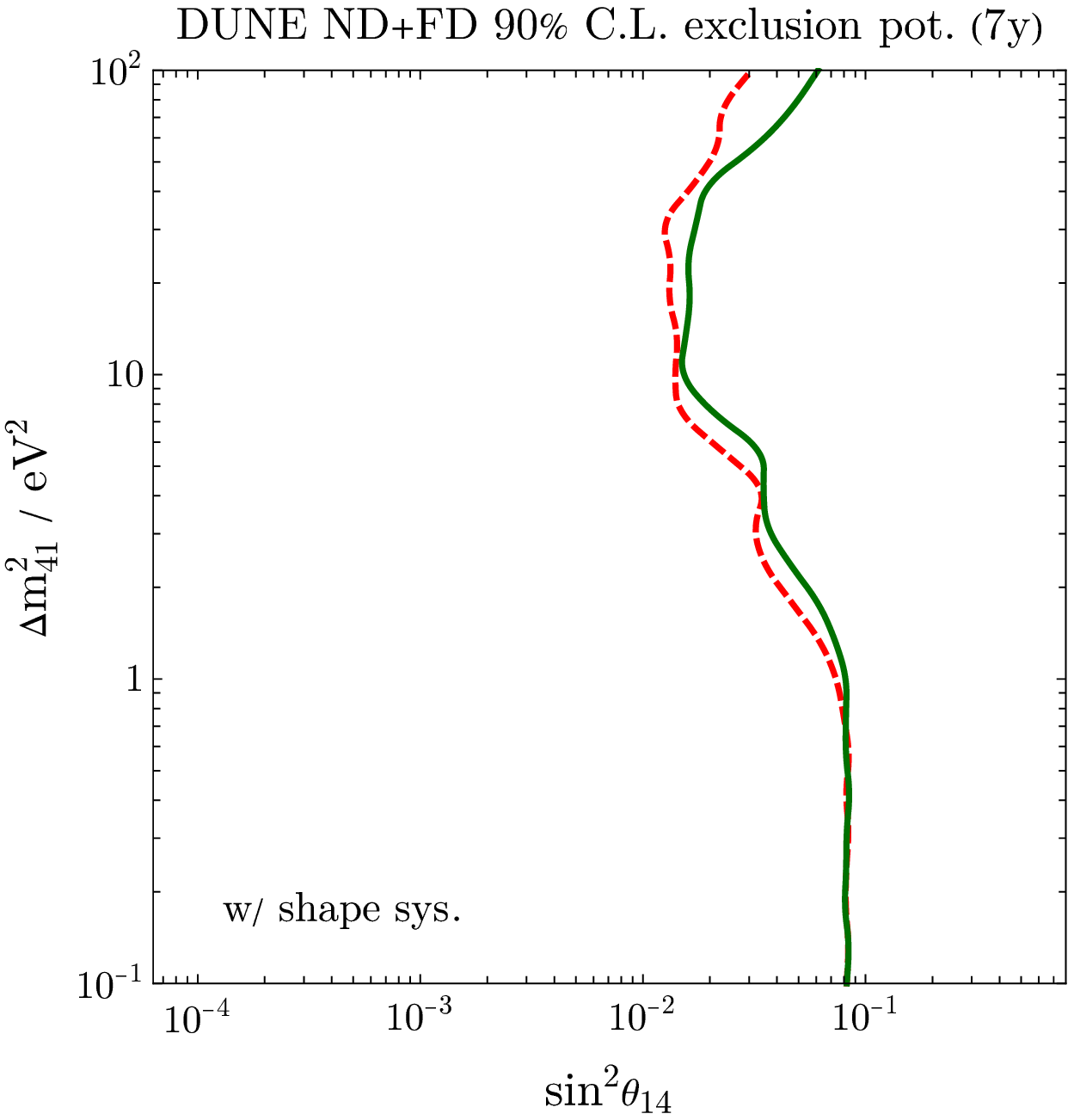}\quad
\includegraphics[height=65mm,width=58mm]{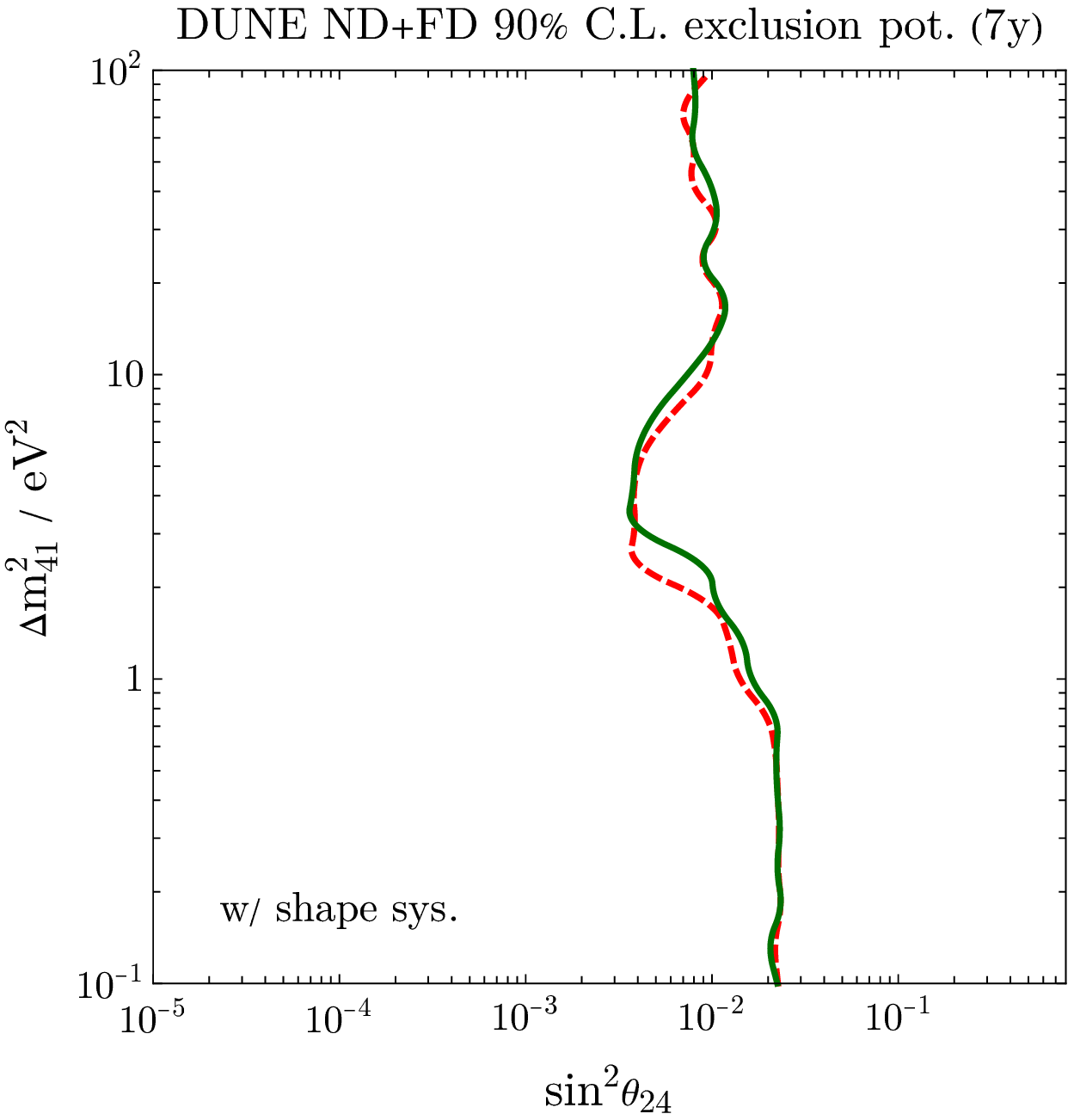}	
\vspace{-0.4cm}
\caption{The same as fig.3 with $\sigma'=5\%$ bin-to-bin uncorrelated systematical errors.}  
\end{figure}
%\vspace{-0.3cm}
Summarizing our main conclusions
\begin{itemize}
\item We considered the effect of a spatial distribution in the oscillation baseline for sterile neutrinos
instead of a single baseline.
%\vspace{-0.3cm}
\item We took into account both DUNE ND and FD event rates with and without energy shape systematics.
%\vspace{-0.3cm}
\item Consequently, DUNE's sterile neutrino exclusion reach is affected with a slight decrease in 
sensitivity, relevant for precision studies.
%\vspace{-0.3cm}
\item This effect is present both with and without energy shape systematics.
\end{itemize}		
%\vspace{-0.6cm}
\newpage
\section*{Acknowledgments}
%\vspace{-0.3cm}
We thank the organizers of Discrete 2022 for providing us with the opportunity to present our results.
We are also indebted to Sampsa Vihonen for extensive support and insighful discussions. J.T.P. further
thanks M. Orcinha for assistance with ROOT. The work of J.T.P. was supported by Funda\c{c}\~{a}o para 
a Ci\^{e}ncia e Tecnologia (FCT, Portugal) through the projects PTDC/FIS-PAR/29436/2017, 
CERN/FIS-PAR/0004/2019 and CFTP Unit 777 (namely UIDB/00777/2020 and UIDP/00777/2020 which are 
partially funded through POCTI (FEDER), COMPETE, QREN and EU. CFTP computing facilities were 
extensively used throughout the project.
% C.R.~Das acknowledges a scholarship
%from the Funda\c{c}\~{a}o para a Ci\^{e}ncia e a Tecnologia (FCT,
%Portugal) (ref. SFRH/BPD/41091/2007), also greatly thanks the Department of Physics,
%Jyv\"askyl\"{a} University, in particular Prof. Jukka Maalampi (HOD) for
%hospitality and financial support.
%This work was partially
%supported by FCT through the projects CERN/FP/123580/2011, 
%PTDC/FIS-NUC/0548/2012 and CFTP-FCT Unit 777 (PEst-OE/FIS/UI0777/2013)
%which are partially funded through POCTI (FEDER).
%\vspace{-0.4cm}
%\newpage

\end{document}